\documentclass[lettersize,journal]{IEEEtran}
\usepackage{amsmath,amsfonts}
\usepackage{algorithmic}
\usepackage{algorithm}
\usepackage{array}
\usepackage[caption=false,font=normalsize,labelfont=sf,textfont=sf]{subfig}
\usepackage{textcomp}
\usepackage{stfloats}
\usepackage{url}
\usepackage{siunitx}
\usepackage{verbatim}
\usepackage{graphicx}
\usepackage{bm}
\usepackage{cite}
\graphicspath{{images/}}
\usepackage{multirow}
\usepackage[dvipsnames]{xcolor}
\newcommand{\blockcomment}[1]{}
\newcommand{\figref}[1]{\figurename~\ref{#1}}
\newcommand{\tabref}[1]{Tab.~\ref{#1}}

\begin{document}
\bstctlcite{IEEEexample:BSTcontrol} 

\title{HARDCORE: H-field and power loss estimation for arbitrary waveforms with residual, dilated convolutional neural networks in ferrite cores} 

\author{\IEEEauthorblockN{Wilhelm Kirchgässner\IEEEauthorrefmark{1}, Nikolas Förster\IEEEauthorrefmark{1}, Till Piepenbrock\IEEEauthorrefmark{1}, Oliver Schweins\IEEEauthorrefmark{1}, Oliver Wallscheid\IEEEauthorrefmark{1}}

\IEEEauthorblockA{\IEEEauthorrefmark{1}Department of Power Electronics and Electrical Drives \\ Paderborn University, 33095 Paderborn, Germany}

}

\maketitle

\begin{abstract}
The MagNet Challenge 2023 calls upon competitors to develop data-driven models for the material-specific, waveform-agnostic estimation of steady-state power losses in toroidal ferrite cores. The following HARDCORE (H-field and power loss estimation for Arbitrary waveforms with Residual, Dilated convolutional neural networks in ferrite COREs) approach shows that a residual convolutional neural network with physics-informed extensions can serve this task efficiently when trained on observational data beforehand. 
One key solution element is an intermediate model layer which first reconstructs the $bh$ curve and then estimates the power losses based on the curve's area rendering the proposed topology physically interpretable. 
In addition,  emphasis was placed on expert-based feature engineering and information-rich inputs in order to enable a lean model architecture.
A model is trained from scratch for each material, while the topology remains the same.
A Pareto-style trade-off between model size and estimation accuracy is demonstrated, which yields an optimum at as low as 1755 parameters and down to below 8\,\% for the 95-th percentile of the relative error for the worst-case material with sufficient samples. 
\end{abstract}


\section{Introduction}
\blockcomment{ 
    \IEEEPARstart{V}{irtually} all electronics depend to some extent on magnetic circuits -- be it in consumer hardware, like mobile applications, or industrial power sources.
    The ever-increasing miniaturization of electronic components and chips on system boards renders thermal and electromagnetic aspects an emerging challenge. 
    A high precision in manufacturing and layout during the design phase becomes more and more urgent, but the mathematical modeling foundation behind power magnetics lacks rigorousness.
    The common practice is, thus, marked by overdimensioning and large material margins for meeting the minimal requirements reliably.
    In order to overcome these adverse circumstances, the MagNet challenge calls for innovative methods that estimate the power loss in an array of diverse scenarios characterized by variations of temperature, frequency, and signal waveform for the magnetic flux and strength in ferrites of a constant toroidal shape.
}
The MagNet Challenge 2023 is tackled with a material-agnostic residual convolutional neural network (CNN) topology with physics-informed extensions in order to leverage domain knowledge. 
Topological design decisions are dictated by peculiarities found in the data sets and by the overall goal of maximum estimation accuracy at minimum model sizes.

\begin{figure}[h]
    \centering
    \def\svgwidth{\linewidth}
    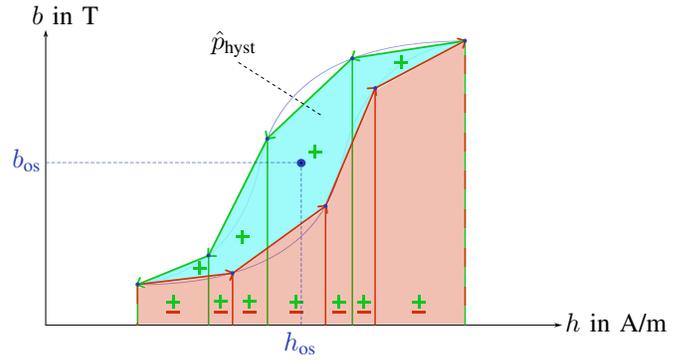
    \caption{Visualization of the shoelace formula applied to a $bh$ polygon.}
    \label{fig:bh_shoelace}
\end{figure}

The topology's central idea is the calculation of the area within the $bh$ polygon based on a preceding $\bm h$ sequence estimate, see \figref{fig:Model_idea_simple}. 
The area within the polygon formed by the sequences $\bm b, \bm h \in \mathbb{R}^{1024}$ can be calculated using the shoelace formula or surveyor's area formula \cite{braden1986surveyor}. 
The shoelace method assigns a trapezoid to each edge of the polgyon as depicted in \figref{fig:bh_shoelace}.
The area of these trapezoids is defined according to shoelace either with a positive or negative sign, according to the hysteresis direction. The negative areas compensate for the parts of positive trapezoids that extend beyond the boundaries of the polygon. 
Provided that the polygon is shifted into the first quadrant by some offsets $h_\text{os}$ and $b_\text{os}$, the power loss in \si{\watt\per\cubic\meter} caused by magnetic hysteresis effects can be computed with the frequency $f$, $M = 1024$, and circular padding by
\begin{align}
    \hat{p}_\text{hyst} &= f \cdot \frac{1}{2} \sum_{i=0}^{M-1} b_{i}(h_{i-1}-h_{i+1}).
    \label{eq:shoelace}
\end{align}


\begin{figure}[htb]
    \centering
    \includegraphics[width=\linewidth]{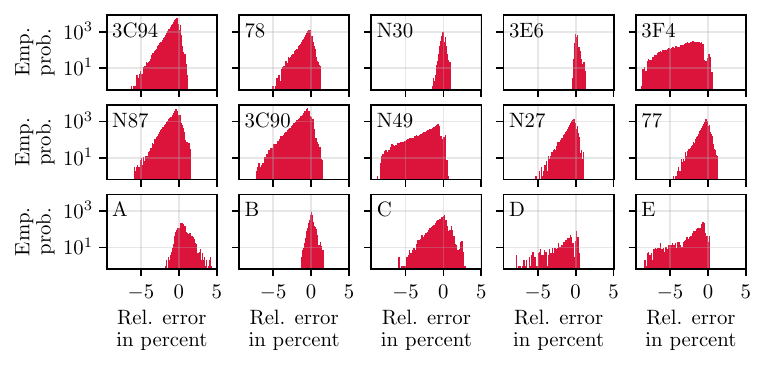}
    \caption{Relative error $(\hat{p}_\text{hyst} - p) / p$ histogram between provided scalar $p$ and $\hat{p}_\text{hyst}$ calculated from the likewise provided $bh$ polygon area.}
    \label{fig:bh_poly_area_err}
\end{figure}

\begin{figure}[htb]
    \centering
    \includegraphics[width=\columnwidth]{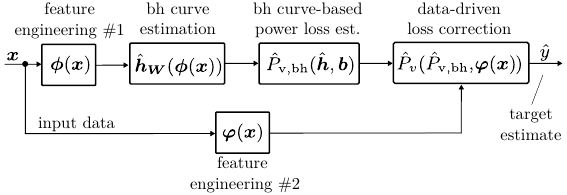}
    \caption{Overview of the physics-inspired HARDCORE modeling toolchain.}
    \label{fig:Model_idea_simple}
\end{figure}

When applying \eqref{eq:shoelace} on the given sequences $\bm b, \bm h \in \mathbb{R}^{M}$ with $M = 1024$, it becomes evident that the calculated area does not equal the provided loss measurements exactly.
\figref{fig:bh_poly_area_err} shows the discrepancy with respect to the provided scalar loss $p$ for all materials.
The relative error ranges up to over 7\,\% for certain materials (e.g. 3F4, N49, D, E).
Consequently, if merely an $h$-predicting model was to be identified, the lower bound on the rel. error would be significantly elevated by this circumstance alone.

Since the power losses calculated from neither the ground truth $bh$ curve area (assuming ideal knowledge on the $\bm{h}$ sequence) nor the estimated area ($\hat{\bm{h}}$ reconstructed via a CNN) do perfectly match the provided loss measurement values (targets), an additional residual correction mechanism is added to compensate for this. 
A high-level view on the proposed residual, physics-inspired modeling toolchain is depicted in \figref{fig:Model_idea_simple}, which is coined the HARDCORE approach (H-field and power loss estimation for Arbitrary waveforms with Residual, Dilated convolutional neural networks in ferrite COREs), and its details are discussed in the following.  


\section{Model description}\label{sec:model_description}

A residual CNN with physics-informed extensions is utilized for all materials.
Such a CNN is trained for each material from scratch.
Yet, the topology is unaltered across materials, signal waveforms, or other input data particularities.


\subsection{One-dimensional CNNs for $h$-estimation}\label{sub:one_dim_cnns}
A 1D CNN is the fundamental building block in this contribution, which consists of multiple trainable kernels or filters per layer slided over the multi-dimensional input sequence in order to produce an activation on the following layer \cite{KriSu2012}.
These activations denote the convolution (more precisely, the cross-correlation) between the learnable kernels and the previous layer's activation (or input sequence).
In this stateless architecture, circular padding ensures that subsequent activation maps are of equal size.
Circular padding can be utilized here instead of the common zero-padding as sequences denote complete periods of the $b$ and $h$ curve during steady state.
Moreover, a kernel does not need to read strictly adjacent samples in a sequence at each point in time, but might use a dilated view, where samples with several samples in between are used.
The dilated, temporal CNN update equation for the $i$-th filter's activation $a^{(l)}_i[k]$ at time $k$ and layer $l$ with the learnable coefficients $\bm W_i \in \mathbb{R}^{A \times \kappa}$ applied on $A$ previous layer's filters, an uneven kernel size of $\kappa \in \{2x+1: x\in \mathbb{N}_0 \}$, and the dilation factor $\delta$ reads
\begin{align}
    a^{(l)}_i[k] &= \sum_{p=0}^{A-1}\sum_{j=-(\kappa-1)/2}^{(\kappa-1)/2} \bm W_{i; (p, j)} \cdot a_p^{(l-1)}[k+j\delta].
    \label{eq:cnn}
\end{align}

Since the task at hand does not require causality of CNN estimates along the time domain (losses are to be estimated from single $\bm{b}$ sequences), the sliding operation can be efficiently parallelized, and sequential processing happens merely along the CNN's depth.
All 1D CNN layers are accompanied by weight normalization \cite{weightnorm2016}. A conceptual representation of the 1D CNN for estimating $\hat{\bm{h}}$ is visualized in \figref{fig:topology} (left part).

\subsection{Feature engineering}
The term feature engineering encompasses all preprocessing, normalization, and derivation of additional features in an observational data set. The input data contains the frequency $f$, the temperature $T$, the measured losses $p$ as well as the 1024 sample points for the $b$ and $h$ waveforms.
Especially the creation of new features that correlate as much as possible with the target variable (here, the $\bm{h}$ curve or the scalar power loss $p$) is an important part of most machine learning (ML) frameworks \cite{Domingo2012}.

\subsubsection{Normalization}
As is typical in neural network training, all input and target features have to be normalized beforehand.
All scalar and time series features are divided by their maximum absolute value that occurs in the material-specific data set, with the exception of the temperature and the frequency, which will be divided by \SI{75}{\celsius} and \SI{150}{\kilo\hertz}, respectively, regardless the material.
Moreover, for an accurate $h$ estimate, it was found to be of paramount importance to normalize each $\bm b$ and $\bm h$ curve again on a per-profile base in dependence not only on the $\ell_\infty$ norm of $|\bm b|$, but also on the maximum absolute $b$ and $h$ appearing in the entire material-specific data set.
The latter two values are denoted $b_\text{lim}$ and $h_\text{lim}$, and can be understood as material-specific scaling constants.
In particular, the per-profile normalized $b$ and $h$ curves for a certain sample read
\begin{equation}
    \bm b_\text{n} = \frac{\bm b}{\max_{k}{|\bm b[k]|}}, \quad \bm h_\text{n} = \frac{\bm h}{h_\text{lim}}\cdot\frac{b_\text{lim}}{\max_{k}{|\bm b[k]|}},
\end{equation}
with $h_\text{lim} = \max_{i,k} |\bm h_i[k]|$, $b_\text{lim} = \max_{i,k} |\bm b_i[k]|$, and $i$ being the sample index in the entire material-specific data set.
Then, $\bm b_\text{n}$ is added to the set of input time series features, and $\bm h_\text{n}$ is the target variable for the $h$ estimation task.

\begin{figure}[h]
    \centering
    \includegraphics[width=\linewidth]{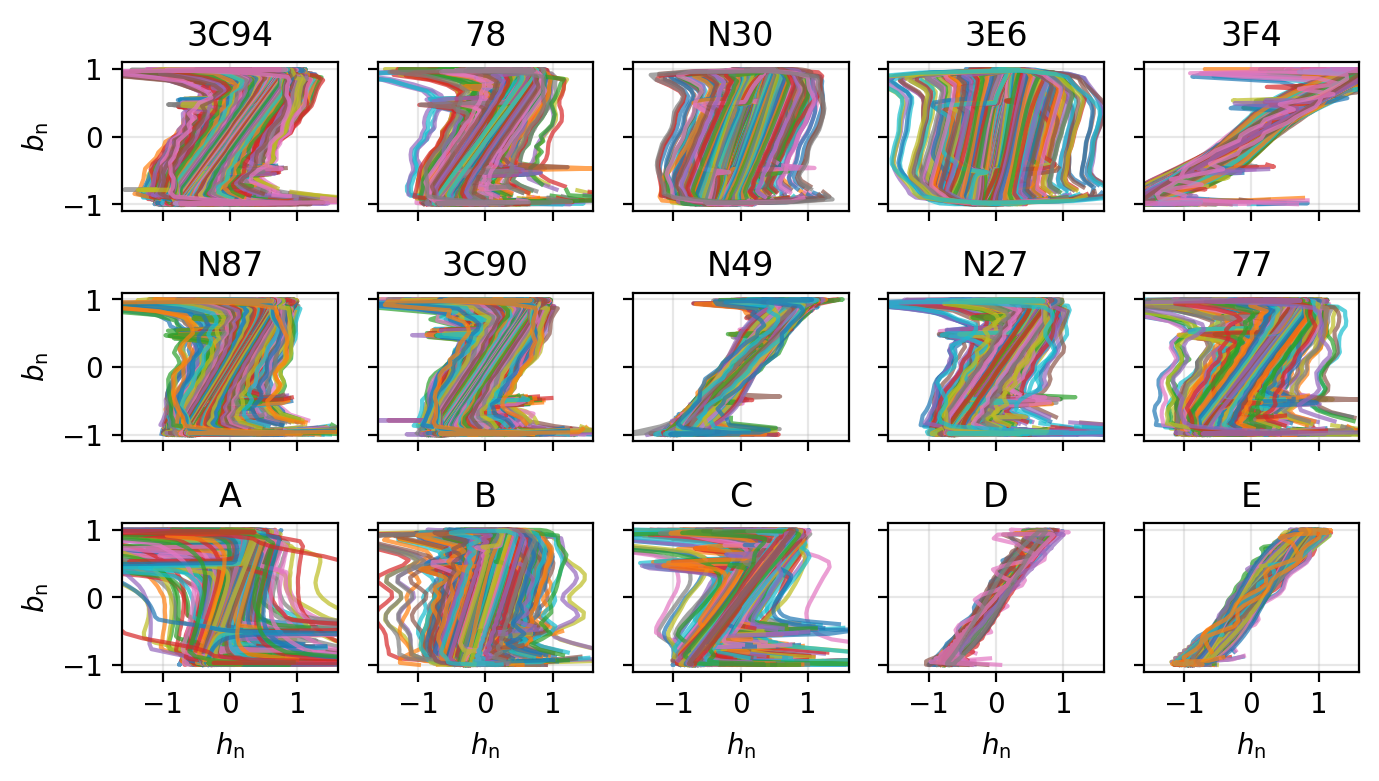}
    \caption{Exemplary samples of the normalized $b$ and $h$ curves.}
    \label{fig:per-profile-norm}
\end{figure}

The $\bm b_\text{n}$ over $\bm h_\text{n}$ curves are displayed in \figref{fig:per-profile-norm}, which underlines how the polygon area becomes roughly unified (no large area difference between samples).
In the following, all features that get in touch with the model are normalized values without any further notational indication.

\subsubsection{Time series features (feature engineering \#1)}\label{subsub:time_series_features}
\begin{figure}[!t]
    \centering
    \includegraphics[width=0.9\linewidth]{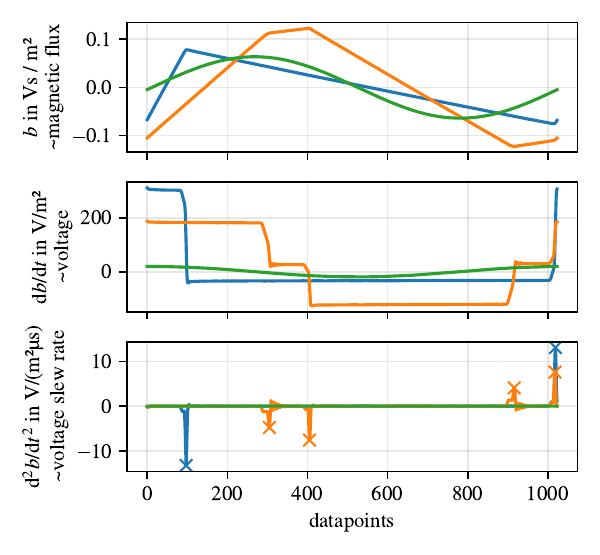}
    \caption{Magnetic flux density examples and their first and second order derivatives for a sinusoidal, triangular and one unclassified waveform with a circuit-based interpretation in terms of their proportionality to magnetic flux, voltage and the voltage slew rate.}
    \label{fig:second_derivative}
\end{figure}

As discussed in Sec.~\ref{sub:one_dim_cnns}, \text{1D}~CNNs build the core of the implemented model.
The inputs to the CNNs are the (per-profile) normalized magnetic flux density $b_\text{n}$ and the corresponding first and second order derivatives ($\dot{b}_\text{n}$ and $\ddot{b}_\text{n}$) as time series.
In a macroscopic measurement circuit context, $\dot{b}$ corresponds to the applied magnetizing voltage throughout the measurement process of the data.
Accordingly, $\ddot{b}$ represents the voltage slew rate during the commutation of the switches in the test setup.
Consequently, the second derivative allows to detect switching events and to characterize them according to their maximum slew rate. 
\figref{fig:second_derivative} shows, that the sinusoidal waveform (green) is generated without any fast transient switching behaviour, probably with a linear signal source.
The nonsinusoidal examples show typical switching behaviour with different voltage slew rates during the single transitions and voltage overshoots as well as ringing.
The second derivative of $\bm b$ informs the ML model about switching transition events and how fast changes in time are.

\subsubsection{Scalar features (feature engineering \#2)}\label{subsub:scalar_features}
Although sequence-based CNNs take up the main share of the ML model size, scalar environmental variables also have a considerable impact on $\bm h$ and $p$. 
While the temperature $T$ is passed to the model unaltered (but normalized), the frequency is presented by its logarithm $\mathrm{ln}(f)$. The sample time $1/f$ is passed directly to the model.
Furthermore, some $b$-derived scalar features are also passed to the model to feed in a priori knowledge.
For example, the peak-to-peak magnetic flux $\Delta b$ as well as the mean absolute time derivative $\overline{|\dot{b}|}$ are directly fed into the network. 
Each waveform is automatically classified into "sine", "triangular", "trapezoidal", and "other" by consulting the form and crest factors, as well as some Fourier coefficients.
The waveform classification is presented to the model by one hot encoding (OHE). 
A summary of all expert-driven input features is presented in \tabref{tab:features}.

\begin{table}[]\centering \caption{Utilized input Features.}
\begin{tabular}{ll|ll}
\multicolumn{2}{l|}{Time series features}                               & \multicolumn{2}{l}{Scalar features}                   \\ \hline
\multicolumn{2}{l|}{}                                                   & \multicolumn{2}{l}{}                                  \\[-1em]
mag. flux density           & $b$                                       & temperature           & $T$               \\
per-profile norm.     & $b_\mathrm{n}$                            &     sample time           & $1/f$  \\
1st derivative              & $\dot{b_\mathrm{n}}$                                 & log-frequency             & $\mathrm{ln}(f)$  \\
2nd derivative              & $\ddot{b_\mathrm{n}}$                                & peak2peak             & $\Delta b$ \\
tan-tan-b                   & $\mathrm{tan}(0.9\cdot\mathrm{tan}(b_\mathrm{n}))$          & log peak2peak         & $\mathrm{ln}(\Delta b)$  \\
                            &                                           & mean abs dbdt         & $\overline{|\dot{b}|}$  \\
                            &                                           & log mean abs dbdt     & $\mathrm{ln}(\overline{|\dot{b}|})$  \\
                            &                                           & waveform (OHE)        &   
\end{tabular}
\label{tab:features}
\end{table}

\subsection{Residual correction and overall topology}
\begin{figure*}[!t]
    \centering
    \includegraphics[width=0.7\linewidth]{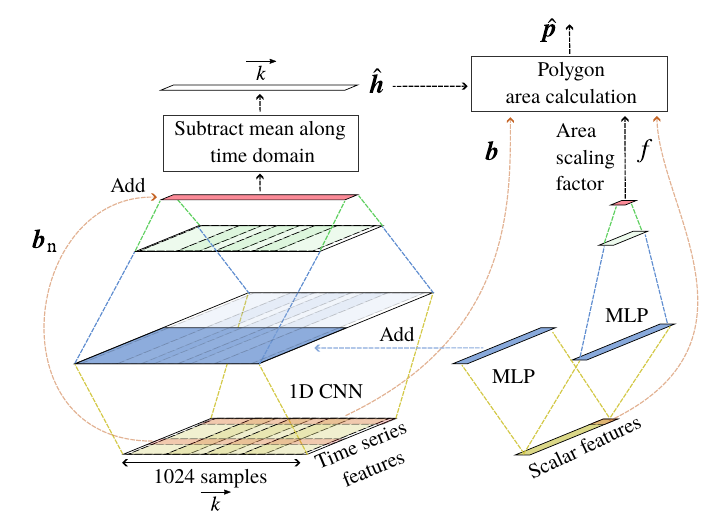}
    \caption{The residual 1D CNN topology is shown while applied on time series and scalar features, which also contain engineered features from \tabref{tab:features}.}
    \label{fig:topology}
\end{figure*}

The model topology comprises multiple branches that end in the scalar power loss estimate $\hat{p}$.
An overview is sketched in \figref{fig:topology}.
Two main branches can be identified: an $h$-predictor and a wrapping $p$-predictor. 
The $h$-predictor utilizes both time series and scalar features with CNNs and multilayer perceptrons (MLPs), and estimates the full $\bm h$ sequence.
The $p$-predictor predicts $p$, on the other hand, and leverages the predicted magnetic field strength $\hat{\bm h}$ with the shoelace formula and a scaling factor.
The latter accounts for the losses inexplicable by the $bh$-curve (recall \figref{fig:bh_poly_area_err}), and is predicted by a MLP that utilizes the scalar feature set only.

The $h$-predictor merges time series and scalar feature information by the broadcasted addition of its MLP output to a part of the first CNN layer output.
This effectively considers the MLP-transformed scalar features as bias term to the time-series-based CNN structure.

On the merged feature set, two further 1D CNN layers follow that end the transformation in a 1024-element sequence.
The per-profile scaled $\bm{b}_\text{n}$ sequence from the set of input time series is element-wise added to this newly obtained estimation (residual connection).
This results in the CNN model to merely learn the difference between $\bm h_\text{n}$ and $\bm{b}_\text{n}$ \cite{BaiKoKo2018}.
Eventually, this sequence becomes the $h$ estimation $\hat{\bm h}_\text{n}$ when the sequence's average along the time domain, that is, across all 1024 elements, is subtracted from each element.
This is a physics-informed intervention in order to ensure a bias-free $\bm h$ estimate $\hat{\bm h}$ after denormalization.
Note that all such operations are still end-to-end differentiable with an automatic differentation framework such as PyTorch \cite{pytorch, GunesBaydin2018}.

Since the resulting $\hat{\bm h}$ can only be trained to be as close as possible to the provided $\bm h$ sequence, which is not leading to the correct $p$ ground truth (cf. \figref{fig:bh_poly_area_err}), another MLP is branched off the scalar input feature set, and denotes the start of the $p$-predictor.
This MLP inherits two hidden layers and concludes with a single output neuron.
This neuron's activation, however, is not $\hat{p}$ but rather an area scaling factor $s \in [-1, 1]$ to be embedded in the shoelace formula \eqref{eq:shoelace} with
\begin{align}
    \hat{p} &= f \cdot \big (0.5 + (0.1 \cdot s) \big ) \sum_{i=0}^{M-1} b_{i}(\hat{h}_{i-1}- \hat{h}_{i+1}).
    \label{eq:shoelace_estimation}
\end{align}
Consequently, the $p$-predictor branch can alter the shoelace formula result by up to $10\,\%$ in positive and negative direction.

The $p$-predictor can be justified physically, when referring to \figref{fig:bh_poly_area_err} again.
As the comparison shows, the hysteresis loss represents the total loss within a variation of $-10~\%$ to $ +5~\%$.
The positive deviation ($\hat{p}_\text{hyst} > p$) indicates some measurement discrepancy between the measured loss and the given $b$ and $h$ curves.
For parts of the negative deviations, a physical explanation can be found in eddy current losses, related to a high dielectric constant and non-zero conductivity.
Due to the small thickness of the used toroidal cores and the limited excitation frequency, eddy current losses are assumed to be of minor effect. 

\blockcomment{ 
\textcolor{red}{The key component of the model is the physically-informed CNN architecture based on the origin of magnetic hysteresis itself, where the current value of $b$ depends on its history.
This can be achieved by CNN applied over a time series. 
Due to the steady-state condition of all given waveforms, circular padding is applied.
This also defines the key benefit over an MLP approach with dense connections, as the weights learned in a CNN are not fix connected to the position of the input waveform $b$.
This means, that the applied CNN architecture is immune against a training bias occurring from the phase angle of the input waveform.
Exemplary this means, that a training with only sines $b$ for the CNN structure does not make any difference compared to cosine.
Differently a MLP network, could try to interpreter the phase of the input signal.
}
}

The physical interpretability of the intermediate estimate $\hat{\bm{h}}$ is a key advantage of the HARDCORE approach:
First, it enables utilizing full $h$ time series simulation frameworks (e.g., time domain FEM solvers).
Secondly, for future designs of magnetic components with arbitrary shapes it becomes indispensable to accurately take into account also geometrical parameters of the core.
This is only possible by distinguishing between the magnetic hysteresis and the (di-)electric losses.

\subsection{Training cost functions}
\begin{figure}[!t]
    \centering
    \includegraphics[width=\linewidth]{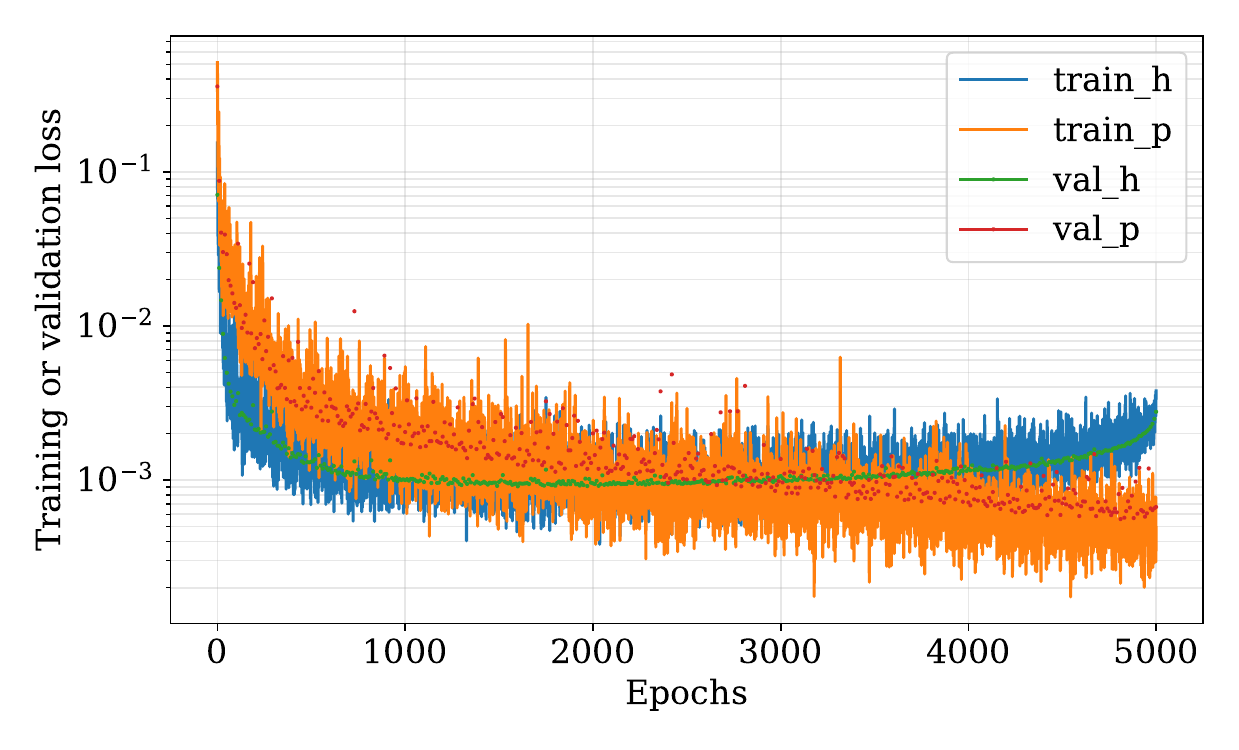}
    \caption{Exemplary training and validation loss curve for material A, seed 0 and fold 3.}
    \label{fig:LearnCurve_Exp_single_58ab3_Material_A_Seed_0_Fold_3} 
\end{figure}

The training process involves two cost functions for a training data set with size $N$: 
First, the $h$ estimation accuracy, which is assessed with the mean squared error (MSE) as
\begin{equation}
    \mathcal{L}_\text{MSE,H} = \frac{1}{NM} \sum^{N-1}_{n=0} \sum^{M-1}_{i=0} (\hat{h}_{i,n} - h_{i,n})^2.
\end{equation}
Second, the power loss estimation accuracy is to be gauged.
Despite the relative error being the competition's evaluation metric, the mean squared logarithmic error (MSLE) is selected
\begin{equation}
    \mathcal{L}_\text{MSLE,P} = \frac{1}{N} \sum^{N-1}_{n=0} (\ln{\hat{p}_n} - \ln{p_n})^2
\end{equation}
in order to not overemphasize samples with a relatively low power loss \cite{tofallis2015better}.
As $\mathcal{L}_\text{MSLE,P}$ also depends on $\hat{\bm h}$ through \eqref{eq:shoelace_estimation}, the question arises, how both cost functions are to be weighted.
In this contribution, a scheduled weighting is applied with 
\begin{equation}
    \mathcal{L}_\text{total} = \alpha\mathcal{L}_\text{MSLE,P} + (1-\alpha)\mathcal{L}_\text{MSE,H},
\end{equation}
where $\alpha = (\beta \cdot i_\text{epoch}) / K_\text{epoch}$ with $\beta \in [0,1]$ denoting a hyperparameter scaling factor, $K_\text{epoch}$ being the number of training epochs, and $i_\text{epoch} \in \{0, 1, \dots K_\text{epoch} - 1\}$ representing the current epoch index.
The scheduled weighting ensures that the model focuses on $\hat{\bm{h}}$ in the beginning of the training, where more information is available.
Later though, the model shall draw most of its attention to the power loss estimate, possibly at the expense of the $h$ estimation accuracy. 
A training example for material A is depicted in \figref{fig:LearnCurve_Exp_single_58ab3_Material_A_Seed_0_Fold_3}.

\section{Hyperparameters, Pareto front and results}
\begin{figure}[!t]
    \centering
    \includegraphics[width=\linewidth]{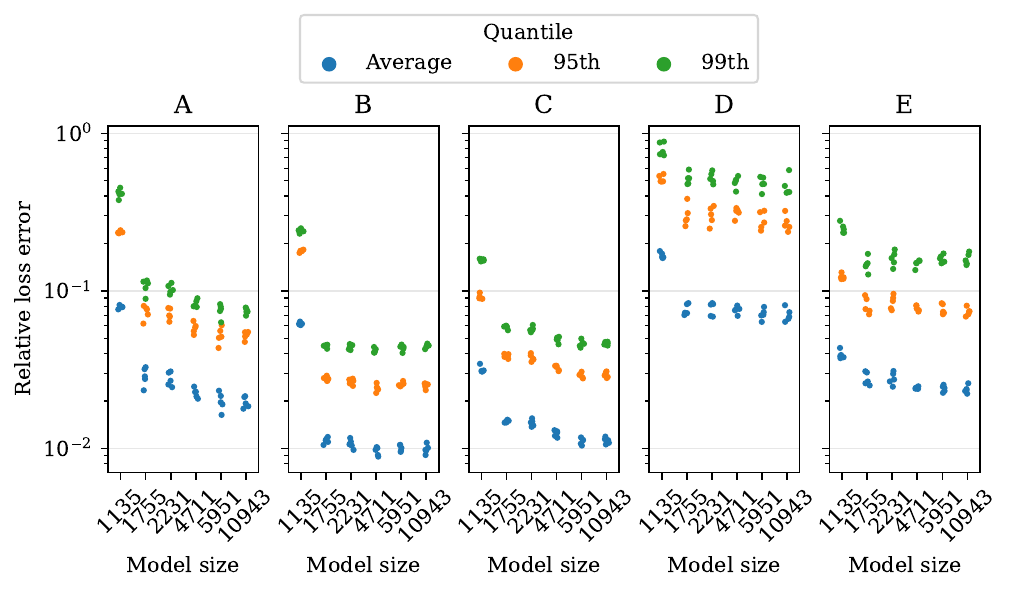}
    \caption{Pareto front for the evaluation materials (A, B, C, D and E) showing model size (amount of parameters) vs. relative error of the power loss estimation.}
    \label{fig:pareto}
\end{figure}

The proposed topology features several degrees of freedom in form of hyperparameters.
An important aspect is the model size, which is defined by the number of hidden layers and neurons in each layer.
A simple trial-and-error investigation can provide fast insights into the performance degradation that comes with fewer model parameters.
In \figref{fig:pareto}, several particularly selected model topologies are illustrated against their achieved relative error versus the inherited model size.
The scatter in each quantile is due to different random number generator seeds and folds during a stratified 4-fold cross-validation.
Topology variations are denoted by the amount of neurons in certain hidden layers.
In addition, the largest topology has an increased kernel size with $\kappa = 17$, and the smallest topology has the second CNN hidden layer removed entirely (the green layer in \figref{fig:topology}).

A slight degradation gradient is evident as of 5\,k parameters for materials A and C, whereas for the other materials the trend is visible only when removing the second hidden layer.
Overall, the material performance scales strictly with the amount of training data available.
Since fewer model parameters are a critical aspect, the chosen final model has 1755 parameters, which is at an optimal trade-off point on the Pareto front.

\begin{figure}[!t]
    \centering
    \includegraphics[width=0.8\linewidth]{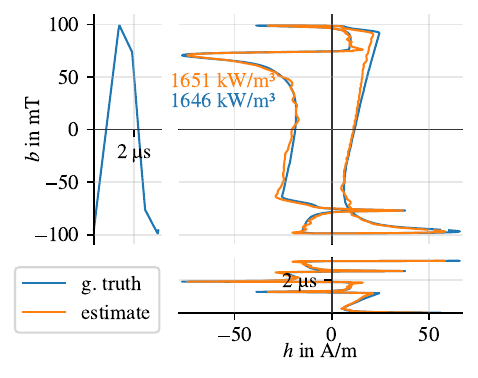}
    \caption{Exemplary ground truth vs.  estimated $bh$ curve comparison.}
    \label{fig:bh_estimation}
\end{figure}

In \tabref{tab:final_model}, the model size of a corresponding PyTorch model file dumped to disk as just-in-time (jit) compilation is reported. 
The exemplary $bh$-curve and $h$-curve estimation is shown in \figref{fig:bh_estimation}.
Reported error rates come from the best seed out of five during a four-fold cross-validation ($\beta = 1, K_\text{epochs} =5000$, Nesterov Adam optimizer).
It shows effectively that any material can be modeled with the same topology at high accuracy as long as a critical training data set size is available (which is not the case for material D, see available training data in \tabref{tab:final_model}).
The final model is already a trade-off between model size and accuracy, such that in case one of the two criteria can be softened, the other can be further improved.

The final model delivery is trained on all training data samples (no repetitions with different seeds), and with $K_\text{epoch}=10000, \delta = 4, \kappa=9$. 
This final topology features a CNN with $12 \text{(TanH)} \rightarrow 8 \text{(TanH)} \rightarrow 1$ \text{(linear)} kernels, a MLP with $11 \text{(TanH)}$ neurons, and a $p$-predictor MLP with $8 \text{(TanH)} \rightarrow 1 \text{(TanH)}$ neurons.

\begin{table}[t]\centering\caption{Final model delivery overview}
\begin{tabular}{c c c c | r r}
    &                &       &       & \multicolumn{2}{c}{Relative error} \\
Material & Parameters & Training  & Model size & Average & 95-th  \\ 
 & & data & & & quantile  \\ \hline
A     & 1755  & 2432            & 43.13 kB      & 2.34\,\% &6.20\,\% \\
B     & 1755  & 7400                & 43.13 kB  & 1.10\,\% & 2.68\,\% \\
C     & 1755  & 5357                & 43.13 kB  & 1.46\,\% & 3.70\,\%  \\
D     & 1755  & 580                & 43.13 kB  & 7.03\,\% &25.76\,\% \\
E     & 1755  & 2013                & 43.13 kB  & 2.51\,\% & 7.10\,\%       
\end{tabular}
\label{tab:final_model}
\end{table}

\section{Conclusion}
A material-agnostic CNN topology for efficient steady-state power loss estimation in ferrite cores is presented.
Since the topology remains unaltered across materials and waveforms at a steadily high accuracy, the proposed model can be considered universally applicable to plenty of materials.
As long as sufficient samples of a material are available (roughly, 2000), the relative error on the 95-th quantile remains below $8\,\%$.
Thus, the contributed method is proposed to become a standard way of training data-driven models for power magnetics.

\bibliographystyle{IEEEtran}
\bibliography{lib}

\blockcomment{ 
    \newpage
    
    \section{Biography Section}
    If you have an EPS/PDF photo (graphicx package needed), extra braces are
     needed around the contents of the optional argument to biography to prevent
     the LaTeX parser from getting confused when it sees the complicated
     $\backslash${\tt{includegraphics}} command within an optional argument. (You can create
     your own custom macro containing the $\backslash${\tt{includegraphics}} command to make things
     simpler here.)
     
    \vspace{11pt}
    
    \bf{If you include a photo:}\vspace{-33pt}
    \begin{IEEEbiography}[{\includegraphics[width=1in,height=1.25in,clip,keepaspectratio]{fig1}}]{Michael Shell}
    Use $\backslash${\tt{begin\{IEEEbiography\}}} and then for the 1st argument use $\backslash${\tt{includegraphics}} to declare and link the author photo.
    Use the author name as the 3rd argument followed by the biography text.
    \end{IEEEbiography}
    
    \vspace{11pt}
    
    \bf{If you will not include a photo:}\vspace{-33pt}
    \begin{IEEEbiographynophoto}{John Doe}
    Use $\backslash${\tt{begin\{IEEEbiographynophoto\}}} and the author name as the argument followed by the biography text.
    \end{IEEEbiographynophoto}

    \vfill
}

\end{document}